\documentstyle[12pt,aaspp,flushrt]{article}
\font\smcap=cmcsc10
\def\la{\mathrel{\hbox{\rlap{\hbox{\lower4pt\hbox{$\sim$}}}\hbox{$<$}}}}
\def\ga{\mathrel{\hbox{\rlap{\hbox{\lower4pt\hbox{$\sim$}}}\hbox{$>$}}}}

\def\Otwo{[O{\smcap ii}]}

\def\etal{{\it et~al.~}}

\def\HST{{\it Hubble Space Telescope}}

\slugcomment{Astrophysical Journal Letters, in press}

\begin{document}

\title{Detailed Mass Map of CL0024+1654 from Strong Lensing}
\author{J. Anthony Tyson, Greg P. Kochanski, and Ian P. Dell'Antonio}
\affil{Bell Laboratories, Lucent Technologies,
700 Mountain Ave., Murray Hill, NJ 07974} 
\affil{Email: tyson@bell-labs.com; gpk@bell-labs.com; dellantonio@bell-labs.com}

\begin{abstract}

We construct a high resolution mass map of
the $z = 0.39$ cluster 0024+1654,
based on parametric inversion  of the
associated gravitational lens.
The lens creates eight well-resolved sub-images of a background galaxy,
seen in deep imaging with the
\HST\footnote{Based on observations with the NASA/ESA \HST, obtained at the
Space Telescope Science Institute, which is operated by AURA, under NASA
contract NAS 5-26555}.
Excluding mass concentrations centered on visible galaxies,
more than 98\% of the remaining mass is represented by a smooth
concentration of dark matter centered near the brightest cluster galaxies,
with a 35 $h^{-1}$ kpc soft core.  
The asymmetry in the mass distribution is $<$3\%
inside $107 ~h^{-1}$\ kpc radius.
The dark matter distribution we observe in CL0024 is far more
smooth, symmetric, and nonsingular than in typical simulated clusters in either $\Omega=1$
or $\Omega=0.3$ CDM cosmologies.
Integrated to $107 ~ h^{-1}$\ kpc radius, the rest-frame mass to light ratio is 
M/$L_V = 276\pm 40~h~(M/L_V)_\odot $.

\end{abstract}


\section{Introduction}

The distribution of mass within high redshift clusters
of galaxies can be a powerful test of theories of gravitational
clustering, and may give clues to the nature of dark matter
(Ostriker \& Cen 1996, Eke \etal\ 1996).  
High-resolution mass maps of clusters would place constraints on both
$\Omega$ and the nature of the dark matter (Crone \etal\ 1994, 1996; Mohr \etal\ 1995) 
because different scenarios for structure formation predict observably different mass 
clumping and segregation in clusters.  
However, it has not previously been possible to map the mass distribution in 
clusters in sufficient detail to view mergers and galaxy-cluster mass segregation.
Neither galaxy light nor hot gas necessarily trace the total mass distribution
precisely (see Schindler \& B\"{o}hringer 1993). 
Although weak lensing, X-ray, and kinematic studies of clusters of 
galaxies set useful limits, there has not been a direct observation on scales 
as small as 1 kpc and $10^9 M_\odot$ of the mass segregation between individual
galaxies and the cluster. Strong lensing of a highly resolved source galaxy can
provide the information required for such a reconstruction.

When multiple gravitationally lensed images
of a complex source occur, the mass density is strongly
constrained because each feature in the source
galaxy must be matched in each of the multiple images.
We construct this high resolution map of the mass distribution in CL0024+1654 
by combining a parametric mass model with a parametric source luminosity model
and making a detailed match to the the HST image.
Qualitative comparisons of this map can be made with 
existing simulations. Low density ($\Omega = 0.3$) flat cosmologies, as well as 
$\Omega = 1$ SCDM, show considerable mass subclustering on 200\ $h^{-1}$ kpc 
scales (Jing \etal\ 1995) for redshifts $z > 0.2$.
Likewise, Frenk \etal\ (1996) separately follow ``galaxies", gas,
and dark matter in SCDM simulations
and find large high contrast lumps of mass surviving to $z = 0.1$ in 
virtually all cases. 
By contrast, simulations of open cosmologies
($\Omega = 0.1, \Lambda = 0$) have very
little substructure even at redshifts of 0.5 and higher
(Jing \etal\ 1995, Evrard \etal\ 1993).
Simulations will soon have sufficient resolution and dynamic range to 
allow quantitative comparison with the data.
Thus, detailed mass maps of clusters
at $z > 0.3$ can help to distinguish between these scenarios.
Here we examine the mass distribution in the central 200 $h^{-1}$ kpc of one such cluster.

\section{Parametric source and mass models}

CL0024+1654 is an optically rich cluster of galaxies at $z ~ = ~ 0.39$
with a velocity dispersion of $\sigma_v = $1200\ km\ s$^{-1}$ 
(Dressler \& Gunn, 1992), and an X-ray luminosity
$L_x=5.0\pm 0.6 \times 10^{43} ~h^{-2}$\ erg cm$^{-2}$\ sec$^{-1}$ (Smail \etal\ 1997 ).
A single background galaxy, easily recognizable 
because of its color and peculiar morphology, is multiply imaged
(see Colley \etal\ 1996).
After subtracting the light of the cluster galaxies, we find a total
of 8 subimages of the background galaxy.
We parametrize the source as 58   
smooth disks of light. Each of these disks is characterized by an
intensity, a scale radius, and the $(x,y)$ position on the source plane (4 parameters).
A source plane resolution of 0.007 arcsec per pixel was chosen
to allow sufficient evaluations of the model to be done within a reasonable time (12 months),
and to allow the model to represent almost all details of the observations.
The disks are overlapping, with a median FWHM of 0.062\arcsec .
Further discussion of 
the morphology of the source galaxy can be found in Tyson \etal\ (1997).
The light from this 
source is then ray-traced through the lens plane, and the resulting image 
is compared on a pixel-by-pixel basis with the HST image.  In this way, we obtain a 
statistically meaningful estimate of the goodness-of-fit.

Because elliptical potentials can be unphysical (Schramm 1994),
we parametrize the mass distribution in CL0024 as a cluster of  
mass concentrations (``mascons").  
Each mascon is based on a power law model (``PL", Schneider et al. 1993)
for the mass density $\Sigma(r)$ vs projected radius $r$,
with both an inner core radius and an outer cutoff radius
\begin{eqnarray}
\Sigma(x) & = &  {K_1  (1 + \gamma x^2) \over (1+x^2)^{2-\gamma}}
 ~~~~~  x < X_o\nonumber\\  
\Sigma(x) & = & K_2 x^{-3} X_o^3  ~~~~~~~~~ x \ge X_o,
\end{eqnarray}

\noindent
where $x = r / r_{\rm core}$, 
$X_o = r_{\rm cutoff} / r_{\rm core}$, and
$\gamma$ is the PL model index. 
The constants $K_2$ and $K_1$ are related by requiring continuity
at $x = X_o $.
${K_1}^{0.5}$ is proportional to the central line-of-sight velocity dispersion.
We build up elliptical mass distributions by superposing
a line of overlapping circular mascons.
In principle, each mascon is described by 9 parameters.
The first four come directly from the equation above
($K_1$,
an inner mass core radius $ r_{\rm core}$,
an outer mass cutoff $ r_{\rm cutoff}$, and
the slope of the mass profile $\gamma$).
For elliptical mass distributions,
there are three
parameters describing the ellipticity 
(the position angle $\theta$, the length of the line of mascons $l_{core}$,
and the uniformity of the spacing of the mascons along the line). 
For mass components not associated with optically observed galaxies,
the $x$ and $y$ positions in the lens plane are also free. 
 
The mass and linear scale sensitivity of this parametric lens inversion
technique vary with position in the cluster; cluster mascons projected
near a long arc have the effect of their mass distribution highly magnified.
For galaxies that are more than about 5\arcsec\ from the arcs,
only their total mass matters, and we parametrize this by the
cutoff radius (because $M \propto r_{\rm cutoff}$).  
Galaxies farther than about  20\arcsec\ from the arcs
are parameterized in groups.
On average, we have one parameter per galaxy.
However, galaxies on the arcs can typically support
several free parameters each.

In practice, one or more mascon
is assigned to each of 118 
cluster galaxies, with the number of free parameters for each mascon 
depending on the distance from the arcs. 
In addition, 25	
free mascons were required for the remaining cluster mass.
We refer to this as the ``dark matter'' [DM], even though it
also includes the mass of the hot X-ray gas.
We will discuss the mass distribution internal to 
the cluster galaxies elsewhere (Dell'Antonio \etal\ 1998).
Two large, diffuse, mascons contain
98\% of the mass not associated with the galaxies.
All parameters were free in both.

In all,
the mass and source models are determined by 512 
parameters.  However,
we have over 3800  
significantly nonzero ($3\sigma$) pixels in the arcs. 
Because the optical point spread function of the WFPC2 is
smaller than one pixel, the signal is nearly uncorrelated even on
neighboring pixels; thus, we have many more independent constraints
than model parameters.
In addition, pixels with no signal serve as constraints,
because they prevent the model from putting flux in areas of the image
where it should not.
The resulting mass model is over-constrained, and is in this sense
unique.  The resolution of the mass model is set by the spacing of the
mass components in the parametric model, and is everywhere much lower than
the formal Nyquist maximum resolution of two pixels.

As we developed the model,
it had enough power to predict the central image,
based on the four major arcs,
then to correctly predict the multiple subimages near the outer arcs.
We measure the fit by taking boxes (approximately 10\arcsec\ square)
around each of the arc images, and creating model images of each arc
using the light and mass models.
The image and models are then compared pixel by pixel.
$\chi^2$ per degree of freedom is then calculated for each realization,
and is iteratively optimized (see Kochanski \etal\ 1998).
Error bars were determined by bootstrap resampling (Efron, 1982) and 
simulated annealing.  For the former, the weight given to each pixel was
varied randomly while allowing the parameters to vary.
In the latter case,
we allow the optimizer to accept increases in $\chi^2$ with
probability $exp( - \Delta \chi^2 )$.
We ran both with the full set of parameters free, and also
with subsets of 7-50 parameters (chosen to capture important correlations) free.
These observed variations in parameter values determine the errors 
(given as 1-$\sigma$ in this Letter),
and control how many parameters may be added
in a given region; too many parameters for a galaxy creates
near-degeneracies and locally large errors.
Our calculated errors naturally include
the effects of correlations between parameters for neighboring mascons.
Nearby galaxies can trade mass, but the total tends to be conserved,
leading to better accuracy on larger scales.

\section{Global mass solution and substructure}

The galaxy masses in the model were initialized using the
Faber-Jackson relation.
When the model evolved to a low $\chi^2$, we performed robustness tests
by perturbing the position and/or mass of a mascon and observing the
reconvergence to the the solution.
Over $2 \times 10^6$ models were searched to reach the solution.
A color rendering of the projected total mass density, including galaxies,
is given in Figure 1.
Optically observed galaxies are fit by small-core PL models, and are 
blue in the figure.
In addition to the large diffuse mascons, other
mascons were added to our fit to allow it to match the
complexity of the cluster's mass distribution.
We will discuss the properties of these small ``dark" objects
in detail elsewhere (Kochanski \etal\ 1998).
Because most of these mascons 
represent small fluctuations in the mass density,
they should generally be regarded
not as distinct objects but as
a representation of the local asymmetry or substructure
in the mass distribution.
Indeed, one of the major results of this Letter is that
these subsidiary mascons are quite light,
and hence the mass distribution is smooth.

The vast majority of the mass is not associated
with the galaxies, and appears as a smooth elliptical distribution
(shown in red in Figure 1) centered near the 
position of the brightest galaxy and elongated in the SE-NW direction.
The elongation is in the same direction as that of the X-ray isophotes.
It is fit by two massive superposed mass distributions
(groups of mascons), 
one with a 34\arcsec\ core radius  
and the other with a 18\arcsec\ core. 
These major mascons are centered within 2\arcsec\ of each other.
This DM not associated with galaxies
shows no evidence of infalling massive clumps:
other than these two major clumps, we find
no dark mascons with 
total mass greater than $5 \times 10^{12} h^{-1} M_\odot$ (1.5\% of 
the cluster mass), out of the 25 in the fit.

Figure 2 shows a contour plot of this dark mass not associated
with galaxies, plotted over the HST blue image.
Figure 3 ({\it top panel}) shows a color image of this diffuse DM,
with the model arcs superposed for reference.
The inner arc (E) may be seen near the center.
The bottom panel of Figure 3 shows the diffuse light component
smoothed with a 0.3\arcsec\  Gaussian,
with the HST images of the arcs superposed in color for reference.
The DM has a soft core,
and is not consistent with the singular mass density profiles
found in cold dark matter simulations (Moore \etal\ 1997; Thomas \etal\ 1997).
We have quantified the strongest allowable singularity
in terms of adding an additional compact mascon into the fit.
We sampled 20 locations for an extra mascon in the central
$30$\ kpc, which would correspond to the difference between a 
Navarro-Frenk-White (1997) [NFW] model and our best-fit profile.
We calculated the fits, allowing correlations between these parameters,
adjacent galaxies, and the main DM clumps.
The largest extra mass the model would support at any
of these locations has a
1-sigma upper limit of $2 \times 10^{11}~ h^{-1} M_\odot$.

The total mass profile is approximately represented by
a single PL model with a
central surface density $7900 \pm 100 ~ h ~ M_\odot ~pc^{-2}$,
a $35\pm 3 ~ h^{-1}$\ kpc core,
and a slope that is slightly shallower than an isothermal
sphere ($\gamma = 0.57\pm 0.02$).
Outside the core, the model is indistinguishable from an
NFW model mass distribution with
$r_{200} = 2500 ~h^{-1}$\ kpc and concentration parameter $c = 8.05$.
However, the presence of a soft core is in disagreement with 
the results from recent CDM simulations.  
For the NFW model which matches the mass distribution outside the core, 
the required mean mass density inside the inner (E) arc's radius is
40\% ($3500 ~h ~M_\odot~ pc^{-2}$) higher than observed (see Figure 4).
This corresponds to an extra interior mass 
of $2\times 10^{12}~ h^{-1}~ M_\odot$,
which we can rule out at $>10\sigma$: 
the position of arc E would be perturbed by over 20 pixels.
Trading mass with the central galaxies also fails.

The total mass enclosed inside the
$ 107 ~ h^{-1}$\ kpc radius of the arcs is 
\begin{equation}
M_{107} = 1.662\pm 0.002 ~ \times 10^{14} h^{-1} d^{-1}_{0.57} ~ M_\odot, 
\end{equation}
where the dimensionless distance ratio 
$d_{0.57}  =  (D_{ls} / D_s) / 0.57  =  1 \pm 0.15$
contains the uncertainty in the source redshift.
The source's featureless spectrum,
star-forming morphology and color,
and presumed \Otwo\ and Ly$_\alpha$ emission lines, suggest
a redshift in the range $1.2 < z < 1.8$.  
Measures of mass segregation and clumpiness and the morphology
are independent of $d_{0.57}$.

To allow a quantitative comparison of our results
with future simulations, we have calculated a clumpiness measure
for the projected mass density $\Sigma(\vec {r})$:
\begin{equation}
C^2 =  A^{-1} \int_{A} \left[{ {\Sigma (\vec {r}) - \Sigma (-\vec {r})} \over 
{\Sigma (\vec {r}) + \Sigma (-\vec {r})} }\right] ^2 d^2r,
\end{equation}
as an average of the normalized density asymmetry over the lens-plane area $A$,
with $\vec {r}$ measured from the centroid that produces the smallest $C$.
This measure is zero for two-fold symmetric mass distributions,
measures the deviation from smoothness,
and is unaffected by ellipticity.

We calculate this statistic several ways:
for mass not associated with luminous galaxies,
and for the total mass distribution,
smoothed on 3 scales.
All the $C$ measures are integrated over a $107~h^{-1}$\ kpc radius circle
centered on the cluster DM ([1950] RA\ =\ 00:23:56.6, DEC\ =\ 16:53:15).
Using Gaussian smoothing with $\sigma$ = 10, 20, and 40 h$^{-1}$ kpc,
we find $C~=~0.071\pm 0.005$, $0.049\pm 0.002$, and $0.036\pm 0.001$,
respectively, 
for the full mass distribution.
If we exclude the galaxies, 
$C~=~0.025\pm 0.003$, $0.029\pm 0.003$, and $0.022\pm 0.002$.
The range of $C$ includes uncertainties in the mass distribution, correction 
factors due to undersampling of the mass distribution, and a 10\% variation 
in the radius of the circle of integration.
This is a very smooth and symmetric distribution,
even with the galaxies included,
and the non-galaxy DM is smoother still.
When comparing the results of N-body simulations with our data using
Equation 3, it is important that the simulations have both sufficient
resolution and enough mass elements to ensure that the simulation's Poisson 
noise does not bias the statistic.

Wilson, Cole, \& Frenk (1996) propose a mass quadrupole measure $Q$(A),
which may also distinguish between clusters in different cosmologies.
For isodensity contours within 10\% of
3820\ $d^{-1}_{0.57}  h ~ M_\odot$ pc$^{-2}$
(which has area of area A\ =\ 1.2\ arcmin$^2$), we measure
$Q$(A)\ =\ $0.028 \pm 0.011 $, for the total mass distribution.
The largest part of the range comes from the choice of contour,
due to the effect of cluster galaxies near the contour.
SCDM simulations typically give $Q$ values over ten times larger than this.

We measure the diffuse intra-cluster light by fitting
a 4-parameter luminosity model to each of the 32 brightest galaxies,
with the innermost 0.5\arcsec\ of each excluded.  
A comparison between luminosity
and mass profiles for galaxies will be given elsewhere (Dell'Antonio
{\it et al.} 1998).
The fit also includes
a sky value and a 7
parameter elliptical PL model for the diffuse cluster light 
($I_o$, $r_{\rm core}$, $\gamma$, $x$,  $y$, $l_{core}$, and $\theta$).
After subtracting the galaxy light,
and excluding disks centered on fainter galaxies,
we find that 15\% $\pm$ 3\%
of the cluster light inside 107\ kpc
radius is diffusely distributed.
This diffuse light, shown in the bottom panel of Figure 3,
has a core size of  7.4\arcsec,
smaller than the DM core size,
but the slope of the distributions are nearly identical
($\gamma_{\rm light} = 0.49$).
The center of the diffuse light
is displaced only 3\arcsec\ WSW from the DM center, and
7\arcsec\ SSW from the total light centroid (which is centered on the large central ellipticals).

To examine the correlation between DM and stellar light we require rest-frame
photometry.
The rest-frame V light is calculated from the observed F814 (791 nm) flux,
since for $z_{cl} = 0.39$ the corresponding emitted wavelength is
$791 /(1 + z_{cl}) = 569$ nm.  Thus, the rest-frame absolute magnitude is
\begin{equation}
M_V = M_{F814} + (M_{V} - M_{569}) - 25 -5~log~D_l -2.5~log(1 + z_{cl}) 
- 2.5~log(F^A_{791}/F^A_{569}),
\end{equation}
where $D_l$ is the luminosity distance, and $F^A$ is the flux of an A$0$ star.
Taking $(M_{V} - M_{569}) = 0.05$ and $2.5~log(F^A_{791}/F^A_{569})= 1.09$, 
$ M_V = -39.83 + m_{F814} + 5~log~h $.
Within a $107 h^{-1}$ kpc radius, the rest-frame V luminosity of the cluster 
is $ 6.0 \pm 0.5 \times 10^{11} ~ h^{-2} L_\odot$, 
yielding a global rest-frame mass-to-light ratio 
of $ 276 \pm 16 ~d^{-1}_{0.57}~h (M/L_V)_\odot $.  
Because of luminosity evolution for $z > 0$ (Kelson \etal\ 1997) this is equivalent
to a zero-redshift $M/L_V$ = $390 ~h~ d^{-1}_{0.57} ~ (M/L_V)_\odot $.

In Figure 4 we plot the radially averaged mass and light profiles for CL0024.
The diffuse component of DM represents about 83\% of the mass 
of the cluster inside 107$h^{-1}$ kpc, while the galaxies contribute the remaining 17\%.
Note that, averaged in this way, the $M/L_V$ ratio increases 30-40\%
from 20\ --\ 100 h$^{-1}$ kpc.

\section{Discussion}

The mass distribution in CL0024 is remarkably relaxed. 
If one assumes Gaussian density fluctuations in an $\Omega=1$ universe,
the fluctuation that seeded CL0024 must have had a very large
amplitude quite early (rare) to have become virialized by $z=0.39$.
One is led to consider non-Gaussian fluctuations or $\Omega << 1$.
In the NFW hierarchical clustering model, our measured
characteristic mass density $\delta_c= 2.6\times 10^4$ implies
(for $\Omega_0=0.3$)
that the 5 largest progenitors contributed 50\% or more of the cluster mass
by a redshift of two (Navarro \etal 1997).
More generally, these observations constrain any theory of structure
formation on 300 kpc scales.

A key result of this first high resolution mass map is the
existence of a 35\ $h^{-1}$\ kpc soft core.
Any possible singularity must be quite small,
contributing less than $2 \times 10^{11} ~ h^{-1} M_\odot$
to the total mass within this core
(10\% of the mass of one of the central elliptical galaxies).
Previous weak and strong lensing studies of clusters have found
evidence for soft mass cores, more compact than the X-ray derived cores
(Tyson \& Fischer 1995; Smail \etal 1996).
Because cold collisionless particles have no characteristic
length scale, the soft core suggests nongravitational interactions.
While HDM can produce soft cores, HDM is not consistent with the
high density of DM that we find in the individual cluster galaxies. 
Because of the relatively low X-ray luminosity (0.7 $L^*_X$), it is very
unlikely that this can be attributed to hot gas alone.
Galaxy halo stripping proceeds more rapidly at larger overdensities 
(Lokas \etal 1996), and the smaller core radius of the diffuse cluster light
is consistent with some baryonic dissipation.

This first high resolution mass map of a cluster of galaxies will be useful to
compare with future N-body/gas-dynamical simulations. 
None of the recent simulations show evidence of a soft core,
in disagreement with these observations. Indeed, as the resolution of
simulations increases from 100 to 30 kpc the central mass becomes more
singular.  Higher resolution gas-dynamical simulations for various initial
fluctuation spectra and values of $\Omega$ will be required.
Comparing with current simulations,
the smoothness of the DM distribution in CL0024+1654
favors open non-flat cosmologies.
Constructing high-resolution mass maps of other clusters will require 
considerably deeper exposures, as we took advantage of an unusual,
complex, high surface brightness galaxy directly behind the cluster.

\medskip
\frenchspacing
\def\refitem{\par\noindent\hangindent 20pt}
\def\AA#1{\it Astron. Astrophys. \bf#1\rm}
\def\ApJ#1{\it Astrophys. J. \bf#1\rm}
\def\AJ#1{\it Astron. J. \bf#1\rm}
\def\MN#1{\it Mon. Not. Roy. Astron. Soc. \bf#1\rm}
\def\UND{\vrule width 1.0in height 0.4pt depth 0pt}
{}

\clearpage
\begin{figure}
\caption{[Color Plate] The reconstructed total mass density in CL0024,
shown as a color-coded mass image.
Dark matter is shown in orange.
Mass associated with visible galaxies is shown in blue.
Contours are at 0.5, 1, and 1.5 times the critical lensing density
(4497\ $h ~ d_{0.57}^{-1} ~ M_\odot$ pc$^{-2}$), with the
heavier contour at the critical lensing density.
This image is 336\ $h^{-1}$\ kpc across. North is up and east is left.
}
\end{figure}

\begin{figure}
\caption{ The reconstructed mass density not associated with visible
galaxies in CL0024 is shown as a contour plot (white contours),
superposed on the F450W (blue) HST image for reference.
Isomass contours for this dark mass are at $0.1 \Sigma_c$ 
intervals in projected mass density,
with thick contour at 1$\times$ $\Sigma_c$, 
as labeled.
The plot is $336~ h^{-1}$\ kpc (100\arcsec\ ) across, centered
at RA\ =\ 00:23:56.6, DEC\ =\ 16:53:15 [1950].
On scales larger than 10 kpc,
this majority component of the DM is remarkably smooth.
Dark matter substructure has already been erased by $z ~=~ 0.39$.
}
\end{figure}

\begin{figure}
\caption{[Color Plate] {\it Top panel}: The cluster mass density not 
associated with galaxies is shown in orange.
The centers of the dark matter (+), total light (x) and diffuse light (o)
distributions are marked.
The reconstructed optical arcs are shown superposed in blue.
{\it Bottom panel}: The diffuse light component of the cluster is shown in 
orange.  Galaxies have been fit and subtracted, and their cores are
replaced with the fit to the diffuse light.
The HST images of the arcs are then superposed in color (one blue and two 
yellow galaxies projected on two of the arcs are reinserted here, 
but are not counted in the diffuse light).
Note the excellent agreement of the modeled arcs ({\it top panel}) 
and the data ({\it bottom panel}).
The demagnified source image just west of the center strongly constrains
the mass profile in the core.
}
\end{figure}

\begin{figure}
\caption{ A radial plot of the mass density and light density.
Total ({\it thick line}) 
and galaxy-only ({\it thin line}) components of the mass are shown. 
The dotted line
is the best NFW fit discussed in the text, and 
the dashed line is the best-fit single PL model.
The 35 $h^{-1}$ kpc soft core in the mass is evident.
A singular mass distribution is ruled out.
The total rest-frame V light profile ({\it solid line}), and galaxy 
V light profile ({\it dashed line}), smoothed with a 5 $h^{-1}$ kpc
Gaussian, are also shown.
}
\end{figure}

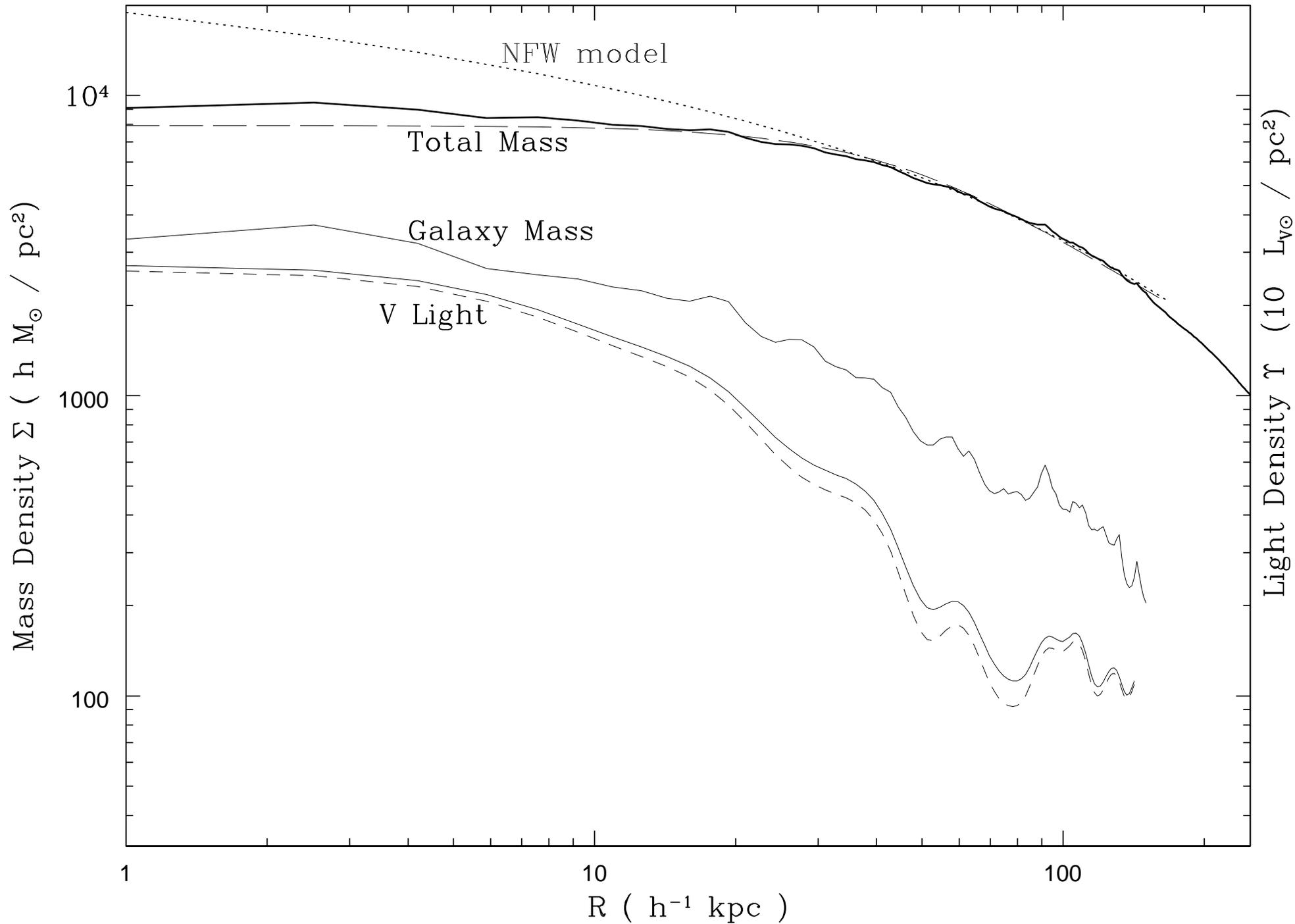
\end{document}